\documentclass[11pt]{article}
\usepackage{geometry}
\usepackage{amsmath}
\usepackage{amssymb,bm}
\usepackage{mathrsfs}
\usepackage{amsfonts}
\usepackage{bm}
\usepackage{graphicx}
\usepackage{endnotes}
\usepackage{array}
\usepackage{sectsty}
\usepackage{url}
\usepackage[multiple]{footmisc}
\usepackage{mathpazo}
\usepackage{quoting}
\usepackage{fnpct}

\newcommand{\gammaminus}{\gamma_{-}}
\newcommand{\gammaplus}{\gamma_{+}}

\usepackage[
    hyperfootnotes=true         
    ]{hyperref}                 

\hypersetup{
colorlinks=true,                
linkcolor=red,                  
anchorcolor=black,              
citecolor=blue,                
urlcolor=cyan,                  
bookmarksnumbered=true,         
linktocpage=false               
}
\begin{document}

\allsectionsfont{\normalsize}
\begin{center}
\large{\textbf{Superluminal Transformations in Spacetimes of Definite Metric}}\footnotemark\footnotetext{This paper was first published in G.\ Hunter, S.\  Jeffers, and J.-P.\ Viger (eds.), \textsl{Causality and Locality in Modern Physics} (Dordrecht:  Kluwer, 1998, pp.\ 227--234).  It was reprinted in Kent A.\ Peacock, \textsl{Quantum Heresies} (London:  College Publications, 2018).  The author thanks Springer Publishing for permission to reproduce the paper in \textsl{Quantum Heresies}.  The publication in \textsl{Quantum Heresies} corrects a sign error that appeared in the original version in Eqs. (12) and (13). }
\end{center}

\begin{center}
Kent A. Peacock\footnote{Department of Philosophy, University of Lethbridge, 4401 University Drive, Lethbridge, Alberta, Canada.  T1K 3M4.  Email:  \url{kent.peacock@uleth.ca}.}
\end{center}

\begin{abstract}
  This paper reviews and extends an
  approach to superluminal
  kinematics set forth by R.\ Sutherland and J.\ Shepanski in 1986.               
  This theory is characterized by  a spacetime with
  positive \emph{definite} metric, 
  and real-valued proper times and proper
  lengths for  superluminal reference frames.  
\end{abstract}

\section{Motivation for this Study}

The central assumption underlying the standard approach to tachyon theory
is that the usual Lorentz transformations apply to the superluminal case.
One therefore simply takes the Lorentz factor $\gamma = \sqrt{1 - \beta^2}$  
(where $\beta \equiv v/c$) and substitutes $\beta > 1$ into it.  This  leads directly
  to imaginary rest masses and proper times for tachyons, with many
attendant  difficulties of interpretation.  (See, e.g., \cite{Feinberg67,Bilaniuk1969}.)

Instead of proceeding by substitution (often a
risky business) it may be useful to attempt to
derive transformations for the superluminal case
from first principles; that is, assume the invariance of the speed of
light and the usual Minkowski spacetime geometry following from that 
postulate, but make the \emph{explicit assumption} that 
it is possible to transform to a superluminal reference frame.
We shall see that  this results in an interestingly different spacetime theory, characterized by a
Lorentz factor of the form $\gamma = \sqrt{\beta^2 - 1}$
and a \emph{definite} metric.

The results derived here were first set forth by Sutherland and 
Shepanski \cite{SS86}, who establish a quite general theory of superluminal reference
frames.  L. Parker \cite{Parker69} also explored a theory with  
definite metric.
The purpose of this note is to draw attention to this approach, and to
present an alternative derivation of Sutherland and Shepanski's
results that indicates in an especially clear way the physical differences between them
and the usual theory.  The definite-metric theory by no means
solves all problems asssociated with the notion of superluminal motion; 
in particular, it does nothing to dispel the closed-loop causal 
paradoxes.  However, in certain
ways it does seem to satisfy the requirements of the Principle of Relativity
in a more natural way than the usual approach.  Furthermore, the theory 
has some very interesting (and indeed pleasing) mathematical properties 
regardless of the question of its physical relevance.

\section{Derivation of Superluminal Transformations}
\subsection{Using Auxiliary Subluminal Frame}
The method used by Sutherland and Shepanski \cite{SS86} involves the use of an
auxiliary subluminal frame. We will not repeat the whole calculation
here.  The essential geometric idea is very natural in the context of Minkowski geometry.
Any boost involves the rotation of
the time axis and the spatial axis in the direction of motion toward
the light cone.  This rotation is symmetric about the light cone---that is,
given a choice of time and distance scales such that the light
cone is at $\pi /4$ with respect to the time and space axes in the 
`lab' frame $S$, both axes rotate toward the light cone through the same
angle.  Now, a superluminal boost will involve the rotation of the time 
and spatial axis (in the direction of motion) \emph{through} the light 
cone; and again, of course, 
this will be symmetrical about the light cone line.  Therefore,
for every (hypothetical) superluminal frame
$\bar{S}$, there exists a subluminal
frame $S'$ with its axes at the \emph{same} angle $\phi$ with respect 
to the axes of the lab frame, but with time and space axes (in the 
direction of motion) interchanged.

Let $\bar{v}$ be the superluminal velocity of $\bar{S}$ with respect 
to the lab frame $S$, and let $v$ be the subluminal velocity of the 
auxiliary frame $S'$ with respect to $S$.  One readily shows that
$\tan \phi = v/c = c/\bar{v}$, giving $v = c^2/{\bar{v}}.$

Let $(x,y,z,t)$ be coordinates in $S$, $(x',y',z',t')$ be the coordinates
in $S'$, and  $(\bar{x},\bar{y},\bar{z},\bar{t})$ be the coordinates in
$\bar{S}$.  Between $S$ and $S'$ there stand the usual subluminal 
Lorentz transformations
\begin{equation}
x' = \gammaminus(x - vt), \;\;
t' = \gammaminus(t - vx/c^2), \;\; y' = y, \;\; z' = z, \label{LT1}
\end{equation}
where we define $\gammaminus \equiv 1/\sqrt{1 - \beta^2}.$
Sutherland and Shepanski show that
by making appropriate substitutions in these formulae, one arrives
at the superluminal transformations
\begin{equation}
\bar{x} = \gammaplus(\bar{v}t - x), \;\;
\bar{t} = \gammaplus(\bar{v}x/c^2 - t) \;\;
\bar{y} = y, \;\; \bar{z} = z. \label{LT2}
\end{equation}
where we define $\gammaplus \equiv 1/\sqrt{\beta^2 - 1}.$
Sutherland and Shepanski simply write the usual $\gamma$ factor with
absolute value  bars, but our notation emphasizes the
physical distinction between the subluminal and superluminal cases.  

\subsection{Using the Galilean Limit}
We now outline an alternative derivation of the superluminal
transformations which makes their physical basis especially
clear.

One familiar way of deriving the subluminal Lorentz transformations is to write down
the transformation rule that would hold for position in the Galilean limit,
and then construct the relativistic picture by assuming that there
is a velocity-dependent correction factor to be determined.  (See, e.g.,
\cite[46--47]{Maudlin94a}.)  We will here apply this method under the 
explicit assumption that the frame to which we transform is moving 
superluminally.

\begin{figure}[h]
\begin{center}
\medskip
\includegraphics[scale=.95]{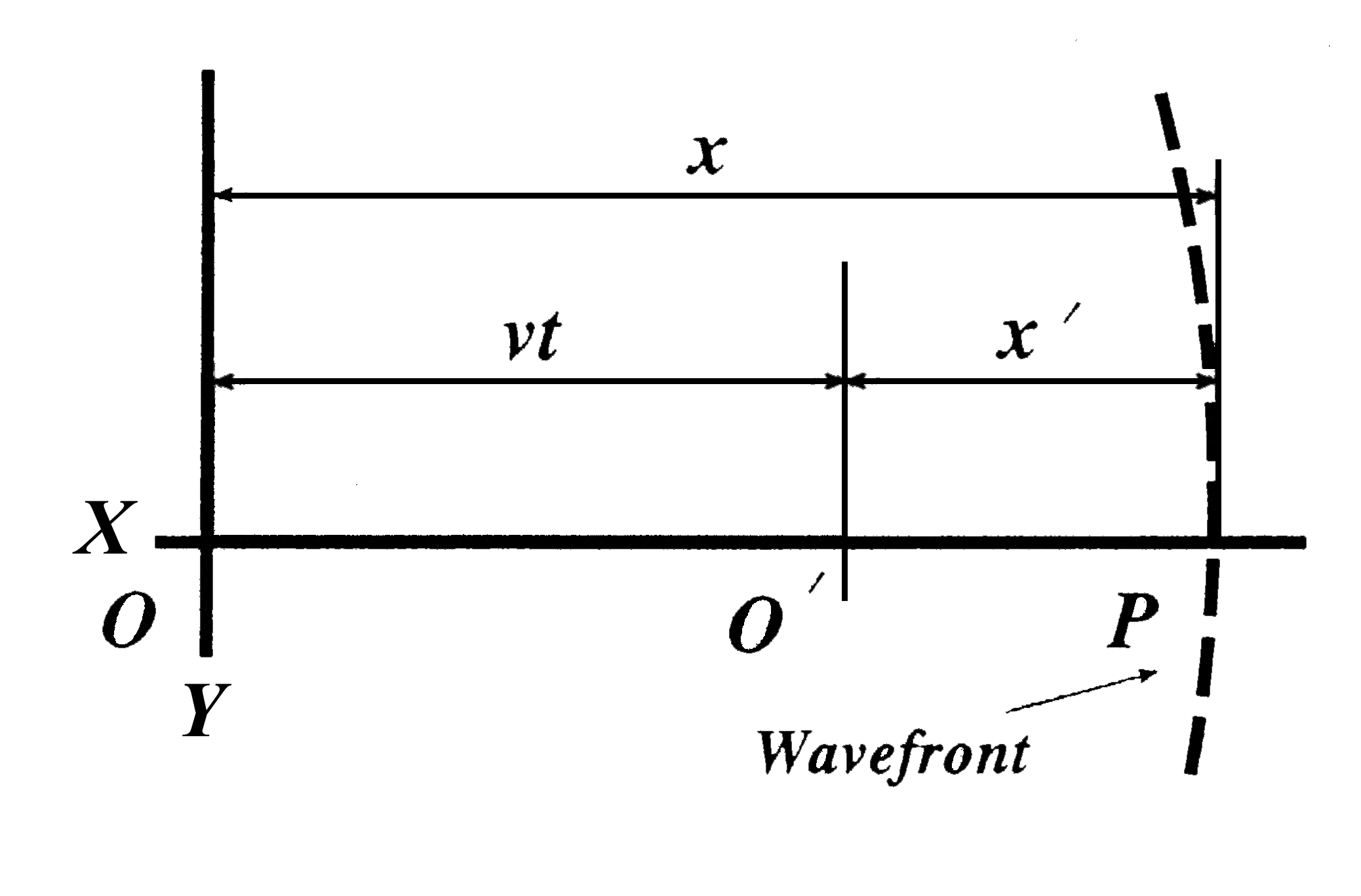} 
 \caption{Subluminal Case}
\bigskip
\end{center}
\end{figure}

\subsubsection{Subluminal Case}
For clarity of comparison, we begin with a 
review of the familiar subluminal derivation.

Our first task is to establish what would hold in the Galilean limit.
Accordingly, we will assume that light moves with some finite velocity
$c$, but we assume Galilean rules for addition of velocities
and the existence of an absolute time.
Now suppose that there are two frames with origins
$O$ and $O'$, with $O$ at rest in the laboratory frame and $O'$ moving along the common $x$-axis with
constant subluminal velocity $v$.  Assume also that a wave-front was emitted from
$O$ at time $t = t' = 0$ and let $P$ be the point where the wave-front cuts the
$x$-axis. Let $x$ be the distance $OP$ in $O$'s coordinates, and $x'$ be
the same distance in $O'$'s coordinates.  As Figure 1 shows, we readily
get \begin{equation}
x'  =  OP - OO'  =  x - vt.
\end{equation}
To get the inverse relationship we note, either from the figure or from
inverting the last equation, that
\begin{equation}
x = x' + vt.
\end{equation}
However, since this is the Galilean picture, the two observers agree on
their time coordinates, and so
\begin{equation}
x = x' + vt'.
\end{equation}

To derive the relativistic transformations we take 
\begin{equation}
x = ct         \;\;
\mbox{and} \;\;  x' = ct'.   \label{Inv}
\end{equation}
These relations express our assumption that $c$ is invariant for both
observers; this is what forces the difference between
the Galilean and relativistic cases.
We also
assume that there is some velocity-dependent correction factor
$\gammaminus$ such that
\begin{equation}
x' = \gammaminus(x - vt),
\end{equation}
with the inverse relationship
\begin{equation}
x = \gammaminus(x' + vt').
\end{equation}
Substituting (\ref{Inv}), we get
\begin{equation}
ct' = \gammaminus t(c - v)  \;\; \mbox{and}  \;\;
ct =  \gammaminus t'(c + v).
\end{equation}
Multiplying the two expressions, we get
\begin{equation}
c^2tt' = {{\gammaminus}^2}tt'(c^2 - v^2);
\end{equation}
i.e.,
\begin{equation}
\gammaminus = 1 / \sqrt{1 - \beta^2}.
\end{equation}
Straightforward substitutions yield (\ref{LT1}).

\subsubsection{Superluminal Case}
We now
make the explicit assumption that the moving system $\bar{O}$ can outrun
the wavefront, and carry out a parallel calculation.
Consider Figure 2, which shows that $\bar{O}$ has \emph{outrun}
$P$.  We let $x$ be the coordinate of $P$ in $O$'s frame and $\bar{x}$ be
its coordinate for $\bar{O}$.  

\begin{figure}[ht]
\begin{center}
\includegraphics[scale=.95]{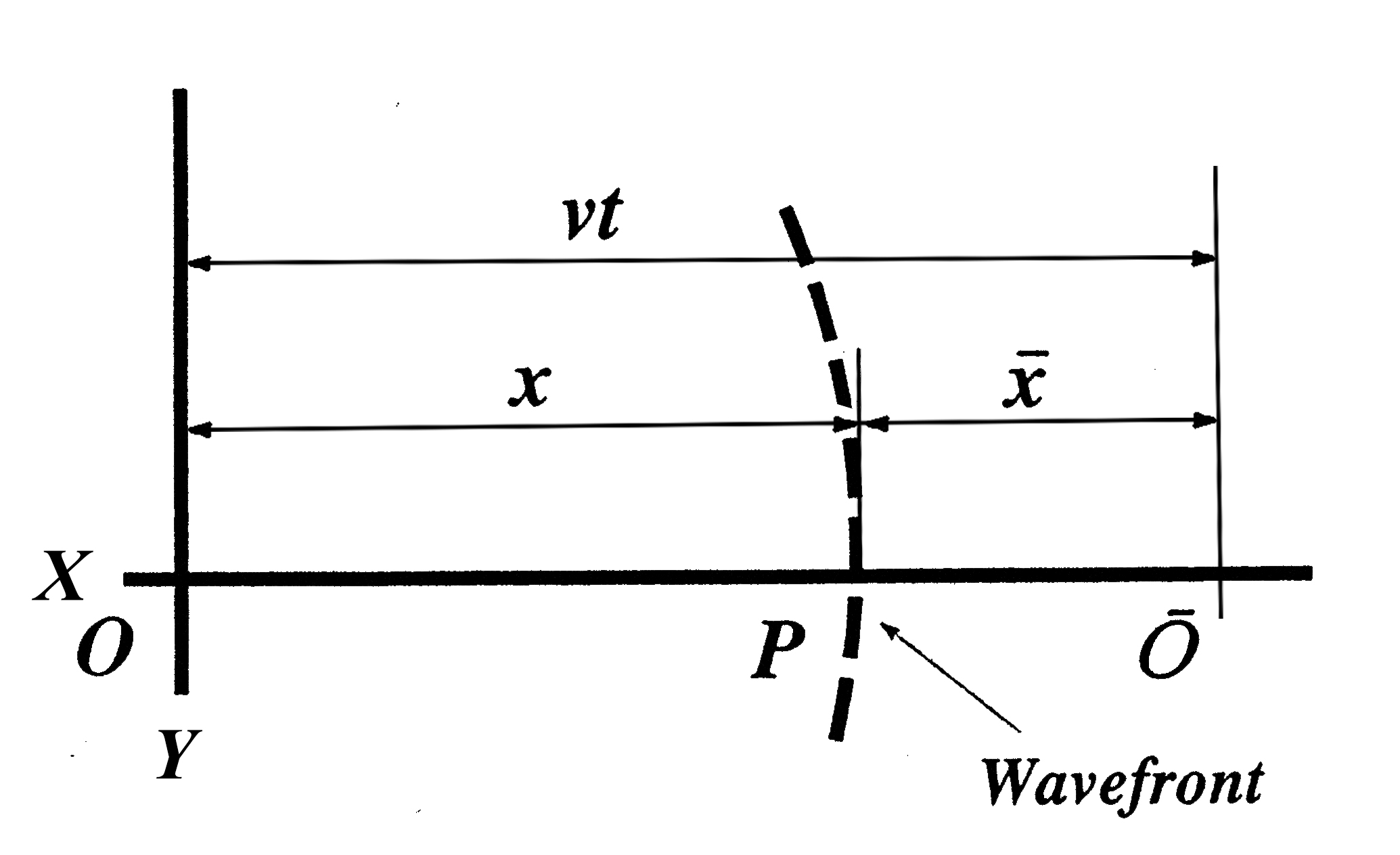} 
 \caption{Superluminal Case}
\bigskip
\end{center}
\end{figure}

Again, we begin with the Galilean limit.  Because $\bar{x}$ must be negative, we have
\begin{equation}
\bar{x} \;  =  \;\; \bar{O}O - OP \;\;  = \; x-vt.
\end{equation}
Inverting, and again noting that the two observers agree that $t = \bar{t}$, we
get the corresponding relationship for $x$:
\begin{equation}
x  =  \bar{x} + vt  =  \bar{x} + v\bar{t}.
\end{equation}

Now we put these transformations for distance to work in order to arrive at
a set of Lorentz-like
transformations for the superluminal case of Figure 2.  As before,
we
assume that there is some velocity-dependent correction factor
$\gammaplus$ such that 
\begin{equation}
\bar{x} = \gammaplus(x - vt),  \label{xbar}
\end{equation}
with the inverse relationship
\begin{equation}
x = \gammaplus(\bar{x}+ v\bar{t}).  \label{xunbar}
\end{equation}
We again take $x = ct$ and  $\bar{x} = c\bar{t}.$
Substituting this condition in (\ref{xbar}) and (\ref{xunbar}), and multiplying as before, we get
\begin{equation}
c^2t\bar{t} = {{\gammaplus}^2}t\bar{t}(v^2 - c^2);
\end{equation}
i.e.,
\begin{equation}
\gammaplus = 1 / \sqrt{\beta^2 - 1}.
\end{equation}
Appropriate substitutions, in this case, yield (\ref{LT2}).
It is therefore crucial to be clear from the outset whether or
not $\bar{O}$ is inside or outside the light cone.

\subsection{Form of the Metric}
Some authors (e.g., Bilaniuk and Sudarshan \cite{Bilaniuk1969}) have defended the 
appearance of imaginary quantities in standard  
tachyon theory by arguing that it merely shows that there is no such thing as being at rest with
respect to a superluminal frame.  However, the Principle of Relativity (PR) implicitly assumes the possibility of
local measurements of positions, times, and masses (which are taken to 
be invariants) for all frames of reference.  Therefore, in order to properly test the PR,
we ought to set up the theory in such a way that these quantities can be real numbers, and the only way to do this 
is to set $ds^2 \ge 0$ everywhere.  
This assumption is implicit in our construction above, since
we take $\bar{x}$, a proper distance in $\bar{S}$, to be
real-valued.  If we could not do this then we would have no
transition to the Galilean limit in the superluminal case,
even though there is no clear reason
why the Galilean limit
(which would treat light like any other disturbance, albeit 
exceptionally fast) should not exist.

As one moves from inside to outside the
light cone, therefore, the signature of the metric must
change, whether expressed in sub- or superluminal coordinates.  
Specifically, time and the distance coordinate in the direction of 
motion must interchange so as to maintain the real-valuedness of 
interval.  For instance, written in subluminal coordinates, the
line element outside the light cone must have the form
\begin{equation}
  ds^2 = c^{2}{dt^2} + dy^2 + dz^2 - dx^2.  \label{FTL_Metric}
\end{equation}
This expresses the fact (evident from inspection of the
Minkowski diagram) that the spatial metric outside the light cone is
hyperbolic, not Euclidean.

A delicate question of interpretation arises.   Sutherland and
Shepanski \cite{SS86} argue that the presence of the geometrically distinguished
spatial direction
indicates that the Principle of Relativity
\emph{cannot} be applied to superluminal frames.  They believe that the 
PR implies that space must be locally isotropic, and therefore locally 
Euclidean.  However, the PR simply requires that there exist a covariant 4-dimensional description of physical
phenomena, consistent with the assumption that the speed of light is an 
invariant.  Nothing suggests that the structure of events cannot look radically different in different 
frames.  Also, there seems to be nothing
in the \emph{General} Principle of Relativity that would prohibit 
locally non-Euclidean frames.  Furthermore, it would be very odd if some 
feature of Minkowski geometry were inconsistent with the PR, since
Minkowski geometry is constructed on the basis of precisely that principle.  Hence, it 
may be that far from ruling out the possibility of tachyons, Sutherland 
and Shepanski's beautiful construction simply gives us (perhaps for the 
first time) an accurate picture of their kinematics. 

It may seem paradoxical to suppose that we can conjoin the assumption of
the invariance of $c$ with the supposition that $\bar{O}$, the origin of 
the moving frame, can be moving faster than light.  The key is that in the superluminal frame 
$\bar{S}$ the wavefront must \emph{recede
with constant velocity $c$}
from any point at rest in 
$\bar{S}$ 
regardless of how fast $\bar{S}$ 
moves with respect to any subluminal frame.  This means that 
(as suggested by (\ref{FTL_Metric})) the 
wavefront in $\bar{S}$ along constant time slices is not a sphere (as it 
must be in subluminal frames) but an hyperboloid of revolution with the axis 
of rotation perpendicular to the direction of superluminal motion \cite{SS86}. 

\section{A Problem for Space Travellers}
The familiar Twin Paradox takes on an interesting twist in the
definite theory.
Suppose there are identical twins Peter and Paul.  Peter remains home on
Earth, while Paul embarks on a subluminal space voyage.  It is well 
known that Paul's elapsed proper time will be less than Peter's; if Paul
travels at relativistic speeds he may even return
home still physiologically young to find his
brother an elderly man. 

Now suppose, \textit{per impossibile} perhaps,
that Paul has the technological means to set out on a \emph{superluminal} voyage.
Let $\beta$ be Paul's velocity (with $\beta \ge 1$),
$t(\beta)$    his elapsed proper time when he returns home,
and $t_0$  Peter's corresponding elapsed proper time.
Then we will have
\begin{equation}
  t(\beta) = t_0\sqrt{\beta^2  - 1}.
\end{equation}
Paul's elapsed proper time is nearly zero when $\beta$
only slightly exceeds 1,
but then begins to increase as
$\beta$ increases, matching Peter's
at $\beta = \sqrt{2}$, and then increasing roughly as $\beta$
thereafter!  If Paul could travel at (say) $10c$, he would age almost 10
times as fast as his brother back on Earth.  Superluminal travel would 
thus offer few advantages to the space traveller.

Space travel enthusiasts (such as this author) may at first find this
result to be discouraging.
However, it might not apply to hypothetical `space warp'
methods of travel \cite{Alcubierre1994}, since conceivably a
locally Euclidean spatial metric could be maintained on board the 
starship.  Of course, this is highly speculative, but it does merit
further investigation.

\section{Causal Paradoxes}
An adequate discussion of causal paradoxes is beyond the scope of this   
paper.  However,
it is easily seen that, \textit{prima facie,} one still gets
closed-loop paradoxes in the
definite theory.  These paradoxes depend upon the topology of the
world-lines, and whether one parametrizes world-lines with real or 
imaginary numbers makes no difference.  
Indeed, as Arntzenius \cite{Arntzenius1994} points out,  
there will be
closed-loop paradoxes in any theory (even a Galilean theory) that allows 
for infinite signal velocities.
The lack of an obvious resolution of the causal
paradoxes in this model should not preclude the
discussion of superluminal frames, however,
because it is essential to explore, in an open-minded fashion,
every avenue that may be mathematically feasible.\footnote{
The author thanks James Robert Brown, B.\ Hepburn,  and participants in the Vigier 
Symposium for useful discussions, and
the University of Lethbridge and the
Social Sciences and Humanities Research Council of
Canada for financial support.}\footnote{Note added in 2023:  since this paper was published in 1998 other papers have
appeared developing superluminal kinematics from a similar perspectives. See, for instance,   R.S.\ Viera, An introduction to the theory of tachyons (`Uma introdu\c{c}\~{a}o \'{a} teoria dos t\'{a}quions')", \textsl{Revista Brasileira de Ensino de  F\'{i}sica} 34(3), 3306--2--3306--15, 2012  (preprint in English at \href{https://arxiv.org/abs/1112.4187}{https://arxiv.org/abs/1112.4187}); J.M.\ Hill and B.J.\ Cox, Einstein's Special Relativity beyond the speed of light, \textsl{Proceedings of the Royal Society A}, 3 October 2012, \href{https://doi.org/10.1098/rspa.2012.0340}{https://doi.org/10.1098/rspa.2012.0340}.   
An important task remaining now is to extend the definite metric approach to dynamics.
}


\end{document}